\documentclass[12pt, a4paper]{article}
\usepackage{amsmath}
\usepackage{amsfonts,amsmath,amssymb,graphicx,hyperref,enumitem,framed,float}
\usepackage{amsfonts,amssymb,amsthm}
\usepackage{color}
\usepackage{siunitx}
\usepackage{booktabs}
\usepackage{algpseudocode,algorithm,algorithmicx}
\usepackage{subcaption}
\usepackage{lscape}
\usepackage{amssymb}
\usepackage{color}
\usepackage{epstopdf}
\usepackage{multirow}
\usepackage{url}
\usepackage[flushleft]{threeparttable} 
\usepackage{setspace}
\usepackage{float} 
\usepackage[round]{natbib}
\usepackage{hyperref}
\usepackage[notablist,nofiglist]{endfloat}
\usepackage{cleveref}
\usepackage[toc,page]{appendix}
\newtheorem{theorem}{Theorem}

\usepackage{graphicx}
\usepackage[top=2.54cm, bottom=2.54cm, left=2.54cm, right=2.54cm]{geometry}

\usepackage[font=small,labelfont=bf]{caption}

\usepackage{pgf}
\usepackage{etex}
\usepackage{tikz,pgfplots}
\pgfplotsset{compat=newest} 
\pgfplotsset{plot coordinates/math parser=false} 
\newlength\figureheight 
\newlength\figurewidth

\algrenewcommand\algorithmicrequire{\textbf{Precondition:}}
\algrenewcommand\algorithmicensure{\textbf{Postcondition:}}
\algdef{SE}[DOWHILE]{Do}{doWhile}{\algorithmicdo}[1]{\algorithmicwhile\ #1}%
\algnewcommand{\algorithmicand}{\textbf{ and }}
\algnewcommand{\algorithmicor}{\textbf{ or }}
\algnewcommand{\Or}{\algorithmicor}
\algnewcommand{\And}{\algorithmicand}
\algnewcommand{\algorithmicgoto}{\textbf{ go to }}
\algnewcommand{\GoTo}{\algorithmicgoto}

\begin{document}


\title{Efficient Pricing of Barrier Options on High Volatility Assets using Subset Simulation\footnote{The authors would like to thank Siu-Kui (Ivan) Au, James Beck, Damiano Brigo, Gianluca Fusai, Steven Kou, Ioannis Kyriakou, Zili Zhu, and the  participants at Caltech, University of Liverpool, and Monash University seminars for helpful comments. Any remaining errors are ours. }}

\author{Keegan Mendonca\footnote{Department of Computing and Mathematical Sciences, California Institute of Technology, E. California Blvd., Pasadena, CA 91125 USA. Email: mendoncakeegan@gmail.com}, Vasileios E. Kontosakos\footnote{Department of Econometrics and Business Statistics, Monash University, Wellington Rd, Clayton, Victoria 3800, Australia. Email: Vasileios.Kontosakos@monash.edu},  Athanasios A. Pantelous\footnote{Department of Econometrics and Business Statistics, Monash University, Wellington Rd, Clayton, Victoria 3800, Australia. Email: Athanasios.Pantelous@monash.edu.}, \\ Konstantin M. Zuev\footnote{Corresponding author. Department of Computing and Mathematical Sciences, California Institute of Technology, E. California Blvd., Pasadena, CA 91125 USA. Email: kostia@caltech.edu}} 

\maketitle
\doublespacing



\abstract
Barrier options are one of the most widely traded exotic options on stock exchanges. 
In this paper, we develop a new stochastic simulation method for pricing barrier options and estimating the corresponding execution probabilities. We show that the proposed method always outperforms the standard Monte Carlo approach and becomes substantially more efficient when the underlying asset has high volatility, while it performs better than multilevel Monte Carlo for special cases of barrier options and underlying assets. These theoretical findings are confirmed by numerous simulation results.




\section{Introduction}
A barrier option is among the most actively-traded path--dependent financial derivatives whose payoff depends on whether the underlying asset has reached or exceeded a predetermined price during the option's contract term \citep{Hull2009,dadachanji2015}. 
A barrier option is typically classified as either \textit{knock -in} or \textit{-out} depending on whether it is activated or expires worthless when the price of the underlying asset crosses a certain level (the barrier) \citep{derman1996ins,derman1997ins,RN48}. Then, the payoff at maturity is identical to that of a plain--vanilla European option, in case the price of the underlying asset has remained above the barrier (for a knock-out barrier option) or zero otherwise. 
Barrier options tend to be cheaper than the corresponding plain vanilla ones 
because they expire more easily and are less likely to be executed \citep{jewitt2015fx}. It was estimated that they accounted for approximately half the volume of all traded exotic options~\citep{RN47}. Despite the 2007--08 credit crunch and the subsequent drop in the demand for path--dependent 
instruments, barrier options can still be a useful investment or hedging vehicle when the structure and the risks of the product are comprehensible.

In the financial industry, barrier options can be traded for a number of reasons, using mostly foreign exchanges, commodities and interest rates as the underlying asset(s). First, barrier options more accurately represent investor's beliefs than the corresponding plain--vanilla options, as a down-and-out barrier call option can serve the same purpose as a plain--vanilla option but at a lower cost, given one has a strong indication that the price of the underlying asset will increase.   
Second, barrier options offer a more attractive risk--reward relation than plain--vanilla options, and their advantage stems from their lower price that reflects the additional risk that the spot price might never reach (knock--in) or cross (knock--out) the barrier throughout its life \citep[further discussion about ins and outs of barriers options can be found in][]{derman1996ins,derman1997ins}. In specific, barrier options on \textit{high} volatility underlying assets can be used in a similar way as cheap deep out--of--the--money options, serving as a hedge to provide insurance in a financial turmoil, given their volatility--dependence \citep{RN42}. 
Hence, the development of a framework able to deal efficiently with barrier options on high volatility underlying assets tackles an actual problem in computational finance, which to our knowledge has not been explicitly studied in past. According to \cite{andersen2001distribution}, 
the mean annualized volatility of the thirty stocks in the \textit{Dow Jones Industrial Average} (DJIA) is approximately equal to 28\% (ranging between 22\% and 42\%) while it is not uncommon to record stocks with volatility levels between 33\% and 40\%. 


Therefore, the pricing of barrier options is a challenging problem due to the need to monitor the price of the underlying asset and compare it against the barriers at multiple discrete points during the contract life \citep{kou2007discrete}. In fact, barrier options pricing provides particular challenges to practitioners in all areas of the financial industry, and across all asset classes. Particularly, the Foreign Exchange options industry has always shown great innovation in this class of products and has committed enormous resources to studying them \citep{dadachanji2015}. However, pricing discretely monitored barrier options is not a trivial task as in essence we have to solve a multi--dimensional integral of normal distribution functionals, where the dimension of the integral is defined by the number of discrete monitoring points \citep{fusai2007analysis}.   

Computationally, certain barrier options such as down-and-out options, can be priced via the standard \textit{Black--Scholes--Merton} (BSM) 
~\citep{RN23}'s paper. This idea can be further extended to more complicated barrier options which can be priced using replicating portfolios of vanilla options in a BSM framework~\citep{RN42}. All these approaches, however, suffer from the BSM model's dependence on a number of assumptions
which are not met in real--world trading \citep{Hull2009}. As a result, the estimates we obtain for option's price under the \textit{equivalent martingale measure} (EMM) are often inaccurate. While there are other models for barrier options with analytical solutions, such as jump-diffusion models \citep{kou2002jump,kou2004option}, the \textit{constant elasticity of variance} (CEV) model \citep{boyle1999pricing,davydov2001pricing}, exact analytical approaches \citep{fusai2006exact}, the Hilbert transform-based \citep{feng2008pricing}, the Laplace transform method built on L\'{e}vy processes \citep{jeannin2010transform} or the Fourier-cosine-based semi-analytical methods \citep{LIAN2017167}, all of them depend on assumptions similar to the ones of the BSM pricing equation. 
Another set of methods for pricing barrier options based on solving \textit{partial differential equations} (PDEs) was proposed in~\cite{boyle1998explicit}, \cite{RN43}, 
\cite{zhu2010fully} and \cite{golbabai2014highly}.  Although these methods are generally powerful, they depend on being able to accurately model the option with PDEs and cannot be used in all circumstances (other approaches used in the pricing of exotic derivatives include the method of lines \citep{chiarella2012evaluation}, where the Greeks are also estimated, robust optimization techniques \citep{bandi2014robust}, applicable also to American options, finite--difference based approaches \citep{wade2007smoothing}, where a Crank--Nicolson smoothing strategy to treat discontinuities in barrier options is presented, and regime--switching models \citep{elliott2014pricing,rambeerich2016high}). As a result, \textit{Monte Carlo simulation} (MCS) is often used for option pricing \citep{schoutens2003pricing} and particularly for barrier options \citep{glasserman2001conditioning}. 

The main advantage of MCS over other pricing methods is its model--free property and its non--dependence on the dimension $N$ of the approximated equation. The latter is an important property since as $N\rightarrow\infty$ ($\Delta t\rightarrow0$), the price of a discretely monitored barrier option converges to that of a continuously monitored one~\citep{broadie1997continuity}. On the other hand, MCS has a serious drawback: it is inefficient in estimating prices of barrier options on \textit{high volatility} assets. Indeed, high volatility makes it difficult for the asset to remain within barriers, which, in turn, makes a positive payoff 
a \textit{rare event} \citep{glasserman1999multilevel}. As a result, any standard MCS method will be inaccurate and highly unstable \citep{geman1996pricing}. This motivates the development of more advanced stochastic simulation methods which inherit the robustness of MCS, and yet are more efficient in estimating barrier option prices. A range of stochastic simulation techniques for speeding up the convergence have been proposed, such as the MCS approximation correction for constant single barrier options \citep{beaglehole1997going},  the simulation method based on the Large Deviations Theory \citep{baldi1999pricing},  and more recently the sequential MCS method \citep{shevchenko2014valuation}. 

The main results of this study can be summarized as follows. First, we develop a novel stochastic simulation method for 
pricing barrier options which is based on the \textit{Subset Simulation} (SubSim) method, a \textit{Markon chain Monte Carlo} (MCMC) --based algorithm originally introduced in \cite{AB01} to deal with complex engineered systems and later extended by \cite{ZWB15} to complex networks \citep[for more details, the reader is referred to][]{AW14}. MCMC provides us with a more efficient way to simulate the quantity of interest, compared to naive MCS methods, by sampling from a target distribution and has been widely used in statistical modelling in finance \citep[see][amongst others for finance--related applications of MCMC]{eraker2001mcmc,philipov2006multivariate,gerlach2011bayesian,stroud2014bayesian}. Here, we apply and further extend this idea to compute both the execution probabilities and prices of barrier options. 

Second, we calculate the fair price for double barrier options on high volatility assets and barriers set near the starting price of the underlying asset. In our framework, the ``failure'' probability corresponds to the probability of the barrier option to be executed at maturity (i.e., the price of the underlying asset to remain withing the barriers). This setting in a simple MCS setup results -- with an extremely large probability -- in asset price trajectories which cross the barriers, rendering the barrier option invalid before maturity. 

Third, we show by measuring the \textit{coefficient of variation} (CV), and the \textit{mean squared error} (MSE) that the proposed SubSim--based algorithm is an efficient technique for the pricing of such derivatives. In particular, the SubSim estimator has a CV which is $O(|\log p_E|^{d/2})$, where $p_E$ is the execution probability and $d \leq 3$ is a constant. Comparing this against the MCS estimator whose CV is $O(p_E^{-1/2})$ and for very small values of $p_E$, we can easily see that the latter increases at a dramatically faster pace compared to the SubSim estimator. Moreover, the MSE of the created SubSim estimator is $O(|\log p_E|^{-k})$ -- where $k \leq 3$ is a constant --,  which decreases for increasing $p_E$. 

Finally, we compare our results against the \textit{Multi--level Monte Carlo} (MLMC) ~\citep{giles2008multilevel,giles2008improved} approach and show that for very small values of the option's survival probability $p_E$ the SubSim estimator outperforms the MLMC estimator in terms of the observed CV. Thus our method can be seen as an alternative to price path--dependent options which also complements MLMC for special cases of underlying assets.       

This paper begins with the introduction of the problem of barrier option pricing and the modification of the SubSim method in order to be able to accommodate it. In section 3, we show how SubSim can be used specifically for the estimation of the execution probability and the option payoff at maturity. Section 4 subsequently presents the main theorem and its proof. This establishes the limiting behaviour of the MSE and the computational complexity for a broad category of applications. Finally, numerical results and comparisons with the standard MCS and the MLMC methods are presented to provide support for the theoretical analysis followed by some concluding remarks. 

\section{Barrier Option Pricing with SubSim}
\subsection{Geometric Brownian Motion (GBM)}
\noindent The starting point in option pricing is modeling the price $S_t$ of the underlying asset. Given the focus of this paper which is more on the simulation and statistical aspects of the method, and less on the modeling of the underlying price process, we use a standard GBM instead of a more complex jump process or a model with stochastic volatility which is frequently used in pricing exotic derivatives \citep[see][amongst others]{kou2002jump,kou2004option, chiarella2012evaluation}. Assume that $S_t$ follows the stochastic differential equation (SDE)
\begin{equation}\label{eq:brown}
dS_t=S_t\mu(t) dt + S_t\sigma(t) dW_t,
\end{equation}
a risk--neutral proces, where $\mu(t)$ is the drift, $\sigma(t)$ is volatility, and $W_t$ is the standard Brownian motion defined on a common probability space $(\Omega, \mathcal{F}, \mathbb{P})$. The discretized solution of (\ref{eq:brown}) can then be written as follows
\begin{equation}\label{eq:asset}
S_n=S_{n-1}\exp\left(\left(\mu_n-\frac{\sigma_n^2}{2}\right)\Delta t+\sigma_n \sqrt{\Delta t}Z_n\right),
\end{equation}
where $Z_1,\ldots,Z_N\sim\mathcal{N}(0,1)$ are i.i.d. standard normal random variables. 

\subsection{SubSim for Barrier Options}
We first consider how SubSim can be used specifically for pricing barrier options and why it is especially efficient for options on assets with high volatility. The goal is to estimate the  barrier option price $P$, which is given by 
the following discounted expectation under the risk--neutral measure $\mathbb{Q}$: 
\begin{equation}\label{eq:price} P= \mathbb{E}\left[h(S_N) \prod_{n=1}^{N} I_{[L_n,U_n]}(S_n) \right],
\end{equation}
where $h(S_N)$ is the payoff at the contract maturity ($t=T$), 
$h(S_N)=\max\{S_N-K,0\}$, $K$ is the strike price, and $I_{[A,B]}(x)$ stands for the indicator function: $I_{[A,B]}(x)=1$ if $A\leq x\leq B$, where $A$ and $B$ are the upper and lower barriers respectively, and zero otherwise. 

In order to use the SubSim method we need to bring the problem in (\ref{eq:price}) in a form suitable to be used as input by the method. 
Suppose that the time--evolution of the dynamic system under study (e.g. evolution of the asset price $S_n$) is modeled by the following discrete model:
\begin{equation}\label{eq:model}
S_n=F(S_{n-1},U_n), \hspace{3mm} n=1,\ldots, N,
\end{equation}
where $S_n$ is the price of the underlying asset at time $t_n$, $S=(S_1,\ldots,S_N)$ is the trajectory of the underlying asset, $U_n$ is a random input at time $t_n$, and $F$ is a certain function that governs the evolution of S (i.e., the GBM (\ref{eq:brown}) in our case).  Let $g(S)$ be the performance function -- a function related to the quantity of interest S -- (e.g. the maximum value of the asset price $g(S)=\max\limits_{n=1,\ldots,N}S_n$). We say that a target event
$E$  occurs if $g(S)$ exceeds a critical threshold $\alpha$:
\begin{equation}\label{eq:event}
E=\left\{ U=(U_1,\ldots,U_N) \hspace{1mm}:\hspace{1mm} g(S(U))\geq\alpha \right\}\subset\mathbb{R}^N. 
\end{equation}

The central idea behind SubSim is to break down the rare event of interest $E$ into a series of ``\textit{less rare}'' events that have easier-to-compute probabilities. This idea is implemented by considering a collection of nested subsets starting from the entire input space $\mathbb{R}^N$ and finishing at the target rare event,
\begin{equation}\label{eq:sequence}
\mathbb{R}^N=E_0\supset E_1\supset \ldots \supset E_L\equiv E.
\end{equation}
The intermediate events $E_i$ can be defined by simply repeatedly relaxing the value of the critical threshold $\alpha$ in (\ref{eq:event}),
\begin{equation} \label{eq:relax}
E_i=\left\{ U=(U_1,\ldots,U_N) \hspace{1mm}:\hspace{1mm} g(S(U))\geq\alpha_i \right\}, \hspace{3mm} \alpha_1<\alpha_2<\ldots<\alpha_L\equiv\alpha. 
\end{equation}

\begin{figure}[h]
	\caption{\textbf{Target event}. The target event $E$ consists of all $Z$-vectors that lead to the positive payoff (option execution). The mapping between $Z$-space and $S$-space is given by (\ref{eq:asset}).}
	\centerline{\includegraphics[width=0.8\linewidth]{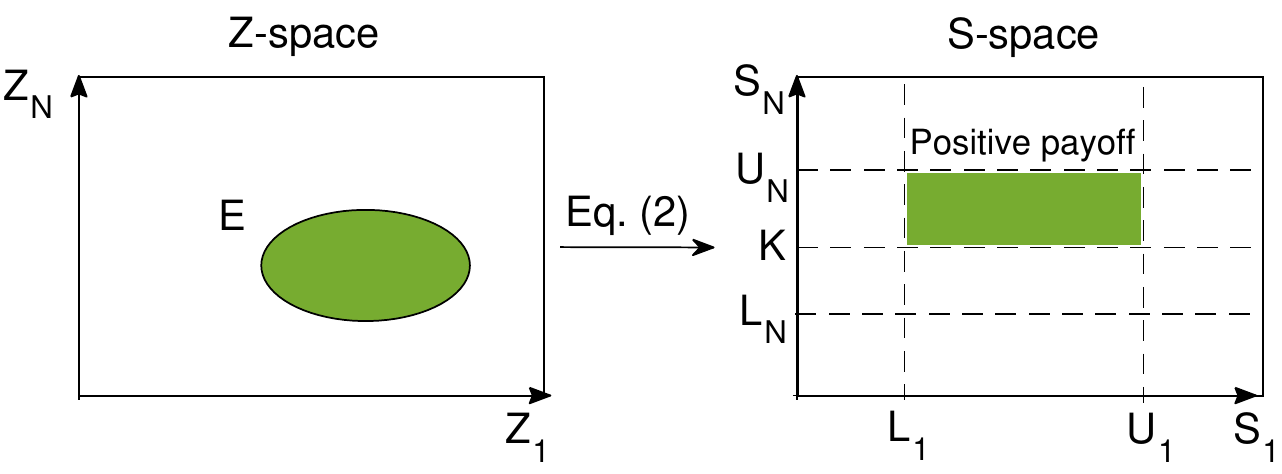}}
	\label{fig3}
\end{figure}

To make SubSim directly applicable, we need to specify suitable functions for the underlying asset price trajectory and the expected payoff at maturity.
Let $E\subset\mathbb{R}^N$ be a set of vectors $Z=(Z_1,\ldots,Z_N)$ that lead to a positive payoff. In other words, $E$ represents the target event for our problem and consists of all vectors $Z$ that result into those asset price trajectories that remain within barriers and end up above the strike price. This is schematically illustrated in Figure~\ref{fig3}.  

Let $\pi$ be the payoff function, 
\begin{equation} \label{eq:payoff}
\pi(Z)=
\begin{cases}
S_N-K, & \mbox{if } Z\in E,\\
0, & \mbox{if } Z\notin E,
\end{cases}
\end{equation}
equal to the payoff of a plain vanilla call in case the asset price trajectory remains within the barriers and ends up above the strike price or zero otherwise. 

\begin{figure}[t]
	\caption{\textbf{Performance function}. The function $g(S)$ quantifies how far the asset price trajectory $S$ is from the positive payoff, which occurs when $S$ stays between the  barriers $U$ and $L$ and ends up above the strike $K$. The value of $g(S)$ on the depicted trajectory is the negative sum of the heights of the vertical bars above the upper barrier (red), below the lower barrier (blue), and ending below the strike (purple). }
	\centerline{\includegraphics[scale = 0.60]{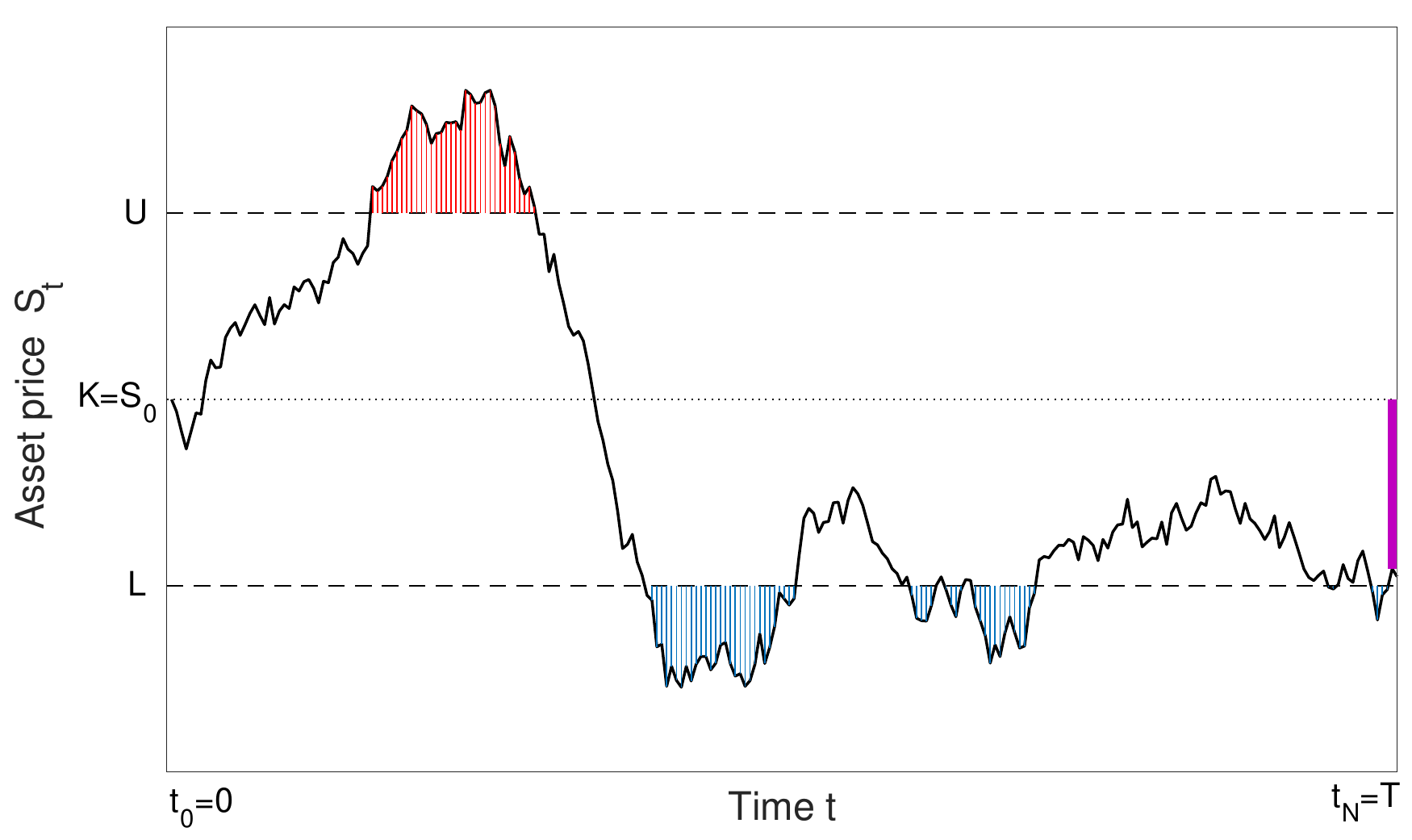}}
	\label{fig4}
\end{figure}

As for the performance function, in the case of option pricing, this quantifies how far the asset price trajectory $S=(S_1,\ldots,S_N)$ lies from the positive payoff, or equivalently, how far $Z=(Z_1,\ldots,Z)$ is from $E$. We define it as follows:
\begin{equation}
g(S)=\sum_{n=1}^N g_n(S_n), 
\end{equation}  
where terms $g_n(S_n)$ quantify how far the asset prices $S_n$ is from the barriers $L_n, U_n$ and strike $K$,
\begin{equation} \label{eq:perffun}
\begin{split}
g_n(S_n)&=\begin{cases}
U_n-S_n, & \mbox{if } S_n>U_n,\\
S_n-L_n, & \mbox{if } S_n<L_n,\\
0, & \mbox{otherwise}.
\end{cases}
\hspace{3mm} \mbox{for }n=1,\ldots,N-1.\\
g_N(S_N)&=\begin{cases}
U_N-S_N, & \mbox{if } S_N>U_N,\\
S_N-K, & \mbox{if } S_n<K,\\
0, & \mbox{otherwise}.
\end{cases}
\end{split}
\end{equation}
The difference between $g_n$ for $n=1,\dots,N-1$ and $g_N$ stems from the fact that at maturity $t_N=T$, the role of the lower barrier is played by the strike price $K$. The performance function $g$ is schematically shown in Figure~\ref{fig4}.  In terms of $g$, the positive-payoff event $E$ can be written, according to the definition of the performance function $g(S)$ in \cref{eq:perffun}, as follows:
\begin{equation}
E=\left\{ Z=(Z_1,\ldots,Z_N) \hspace{1mm}:\hspace{1mm} g(S(Z)) \geq0 \right\},
\end{equation}
where $\alpha$ is now replaced by zero and the defined performance function brings the problem of estimating the probability of positive payoff $p_E$ into the general SubSim framework developed in \cite{AB01}.

Then, combining equations (\ref{eq:payoff}) and (\ref{eq:perffun}), the option price, which in our case is the expected payoff of the contract at maturity, can be rewritten as follows:
\begin{equation}\label{eq:trick}
\begin{split}
P&=\mathbb{E}[\pi(Z)]\\
&=\mathbb{E}[\pi(Z)|Z\in E]\mathbb{P}(Z\in E)+\mathbb{E}[\pi(Z)|Z\notin E]\mathbb{P}(Z\notin E)\\
&=\mathbb{E}[\pi(Z)|Z\in E]\mathbb{P}(Z\in E)=
\mathbb{E}[S_N-K|Z\in E]\mathbb{P}(E)\\
&= \mathbb{P}(E)(\mathbb{E}[S_N|Z\in E]-K).
\end{split}
\end{equation}
Now, the problem boils down to estimating the execution probability $p_E = \mathbb{P}(E)$
and the expectation of the payoff at maturity, given by the second term in the product of \cref{eq:trick}. 

\section{Probability of contract execution $p_E$ and option payoff via SubSim}\label{sec3}

We start with the calculation of $p_E$ to notice that given the sequence (\ref{eq:sequence}), the small probability $p_E$ of rare event $E$ can be written as a product of conditional probabilities:
\begin{equation}\label{eq:decomposition}
\begin{split}
p_E&=\mathbb{P}(E_L)=\mathbb{P}(E_{L}|E_{L-1})\mathbb{P}(E_{L-1})\\&=\mathbb{P}(E_{L}|E_{L-1})\mathbb{P}(E_{L-1}|E_{L-2})\mathbb{P}(E_{L-2})
=\ldots=\prod_{i=1}^L\mathbb{P}(E_i|E_{i-1}). 
\end{split}
\end{equation}
By choosing the intermediate thresholds $\alpha_i$ appropriately (in the  actual implementation of SubSim described below,  $\alpha_i$ are chosen adaptively on the fly), we can make all conditional probabilities $\mathbb{P}(E_i|E_{i-1})$ sufficiently large, and estimate them efficiently by MC-like simulation methods. In fact, the first factor in the right-hand side of (\ref{eq:decomposition}), $\mathbb{P}(E_1|E_{0})=\mathbb{P}(E_1)$,  can be directly estimated by MCS:
\begin{equation}\label{eq:E1}
\mathbb{P}(E_1)\approx \frac{1}{m}\sum_{i=1}^m I_{E_1}\left(U^{(i)}\right), \hspace{3mm} U^{(1)},\ldots,U^{(m)}\sim f_U. 
\end{equation} 

Estimating the remaining factors $\mathbb{P}(E_i|E_{i-1})$ for $i\geq2$ is more difficult since this requires sampling from the conditional distribution $f_U(u|E_{i-1})\propto f_U(u)I_{E_{i-1}}(u)$, which is a nontrivial task, especially at later levels, where $E_{i-1}$ becomes a rare event. In SubSim, this is achieved by using the so-called \textit{modified Metropolis algorithm} (MMA) \citep{AB01,ZK11}, which  belongs to a  large family of MCMC algorithms \citep{Liu01,RC04} for sampling from complex probability distributions. 
The MMA algorithm is a component-wise modification of the original Metropolis algorithm \citep{M53}, which is specifically tailored for sampling in high dimensions, where the original algorithm is known to perform poorly \citep{KZ08}. 

To sample from $f_U(u|E_{i-1})$, MMA generates a Markov chain whose stationary distribution is $f_U(u|E_{i-1})$. The key difference between MMA and the original Metropolis algorithm is how the ``\textit{candidate}'' state of a Markov chain is generated (in \cref{appendix0}, the MMA algorithm used for the sampling is presented). Then, using the detailed balance equation, it can be shown \citep[see][for details]{AB01} that if $U^{(j)}$ is distributed according to the target distribution, $U^{(j)}\sim f_U(u|E_{i-1})$, then so is $U^{(j+1)}$, and $f_U(u|E_{i-1})$ is thus indeed the stationary distribution of the Markov chain generated by MMA. Now, to estimate the small probability of execution $p_E$ the method starts by generating $m$ MCS samples $U^{(1)},\ldots,U^{(m)}\sim f_U$ and computing the corresponding system trajectories $S^{(1)},\ldots,S^{(m)}$ via (\ref{eq:model}) and performance values $g^{(i)}_{U}= g(S^{(i)})$. Without loss of generality, we can assume that
\begin{equation}\label{eq:ordering}
g^{(1)}_{U}\geq g^{(2)}_{U}\geq\ldots\geq g^{(m)}_{U}.
\end{equation} 
Indeed, to achieve this ordering,  we can simply renumber the samples accordingly. Since $E$ is a rare event, all $U^{(i)}\notin E$ with large probability. The ordering (\ref{eq:ordering}) means however that, in the metric induced by the performance function, $U^{(1)}$ is the closest sample to $E$, $U^{(2)}$ is the second closest, etc. Let's define the first intermediate threshold $\alpha_1$ as the average between the performance values of the $\tilde{m}^{\mathrm{th}}$ and $(\tilde{m}+1)^{\mathrm{th}}$ system trajectories, where $\tilde{m}=\beta m$ with $\beta\in(0,1)$:
\begin{equation}
\alpha_1=\frac{g^{(\beta m)}_{U}+g^{(\beta m+1)}_{U}}{2}, \hspace{5mm} 0<\beta<1. 
\end{equation}
Setting $\alpha_1$ to this value has two important corollaries: (1) the MCS estimate of $\mathbb{P}(E_1)$ given by (\ref{eq:E1}) is exactly $\beta$, and (2) samples $U^{(1)},\ldots,U^{(\beta m)}$ are i.i.d. random vectors distributed according to the conditional distribution $f_U(u|E_1)$.

\begin{figure}[t]
	\caption{\textbf{Schematic illustration of Subset Simulation}. First, Monte Carlo samples $U^{(1)},\ldots,U^{(m)}$ are generated. Next, $\tilde{m}=\beta m$ ``seeds'' (the closest samples to $E$) are chosen and MMA is used to generate $V^{(1)},\ldots,V^{(m)}$ from these seeds in the direction of $E$.  The SubSim algorithm proceeds in this way until the target rare event $E$ has been reached and sufficiently sampled. In this visualization, $m=6$ and $\beta=1/3$.}
	\centerline{\includegraphics[width=0.8\linewidth]{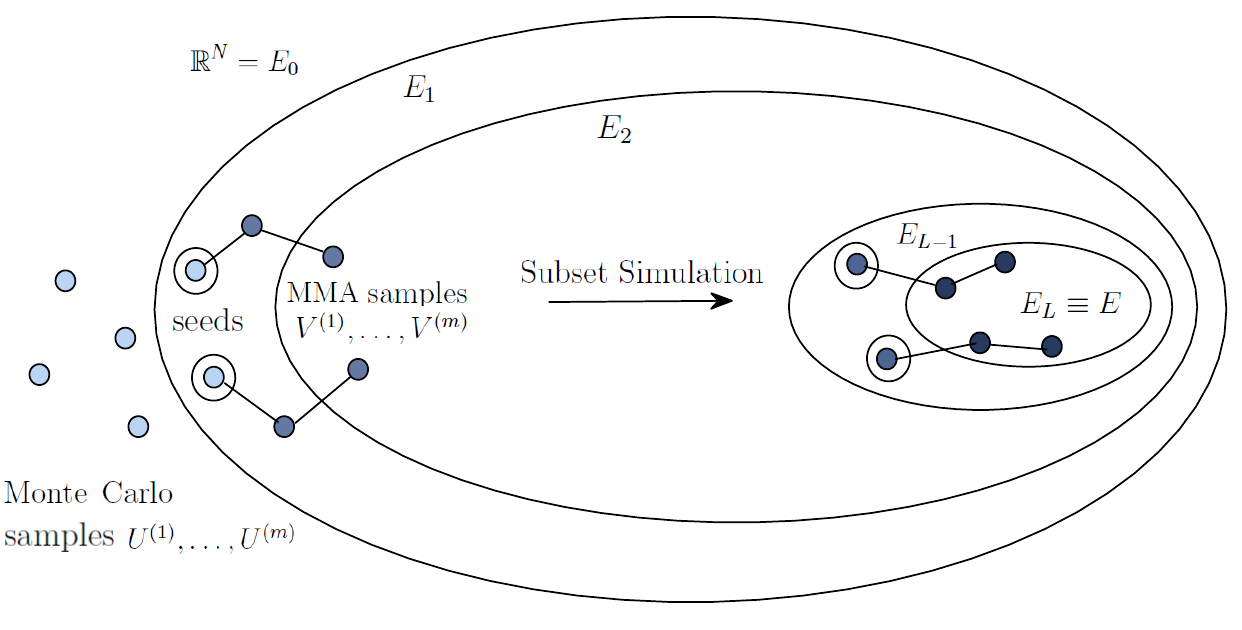}}
	\label{fig2}
\end{figure}

In the next step, SubSim generates $\tilde{m}=\beta m$ Markov chains by MMA starting from $\tilde{m}$ most closest to $E$ samples $U^{(1)},\ldots,U^{(\beta m)}$ as ``\textit{seeds}'':
\begin{equation}
\begin{split}
U^{(i)}&=V^{(i,1)}\stackrel{\mathrm{\footnotesize MMA}}{\longrightarrow}V^{(i,2)}\stackrel{\mathrm{\footnotesize MMA}}{\longrightarrow}\ldots\stackrel{\mathrm{\footnotesize MMA}}{\longrightarrow}V^{(i,l)}.
\end{split}
\end{equation} 
Since by construction, all seeds are in the stationary state, $U^{(i)}\sim f_U(u|E_1), i=1\ldots,\tilde{m}$, so are all Markov chains states $V^{(i,j)}\sim f_U(u|E_1), j=1,\ldots,l$.  The length of each chain is $l=1/\beta$, which makes the total number of states $\tilde{m}l=m$. To simplify the notation, let's denote samples $V^{(i,j)}$ by simply $V^{(1)},\ldots,V^{(m)}$. Next, the second intermediate threshold $\alpha_2$ is  similarly defined as follows: 
\begin{equation}
\alpha_2=\frac{g^{(\beta m)}_{V}+g^{(\beta m+1)}_{V}}{2}, 
\end{equation}
where $g^{(1)}_{V}\geq g^{(2)}_{V}\geq\ldots\geq g^{(m)}_{V}$ are the ordered performance values corresponding to samples $V^{(1)},\ldots,V^{(m)}$. Again, by construction, $\mathbb{P}(E_2|E_1)\approx\beta$ and $V^{(1)},\ldots,V^{(\beta m)}\sim f_U(u|E_2)$. The SubSim method, schematically illustrated in Figure \ref{fig2}, proceeds in this way by directing Markov chains towards the rare event $E$ until it is reached and sufficiently sampled. Specifically, it stops when the number $m_E$ of samples in $E$, which a priori $0\leq m_E\leq m$, is $m_E\geq \beta m$. All but the last factor in the right-hand side of (\ref{eq:decomposition}) are then approximated by $\beta$ and $\mathbb{P}(E|E_{L-1})\approx m_E/m$. This results into the following estimate:
\begin{equation}\label{eq:SSestimate}
p_E\approx \hat{p}_E^{SubSim}=\beta^{L-1}\frac{m_E}{m},
\end{equation} 
where $L$ is the number of subsets in (\ref{eq:decomposition}) required 
to reach $E$. The total number of samples used by SubSim is then
\begin{equation}\label{eq:totalSS}
M=\underbrace{m}_{MCS}+\underbrace{m(1-\beta)(L-1)}_{MMA}.
\end{equation}

The first factor, the probability of positive payoff $p_E=\mathbb{P}(E)$, can be readily estimated by SubSim,
\begin{equation}\label{eq:SSp}
\mathbb{P}(E)\approx \hat{p}_E^{SubSim}.
\end{equation}

Moreover, the conditional expectation in (\ref{eq:trick}) for the terminal asset price can be estimated using the samples generated by SubSim at the last level.  Namely, let $Z^{(1)},\ldots,Z^{(m)}$ be the last batch of MMA samples generated by SubSim before it stops,
\begin{equation}
Z^{(1)},\ldots,Z^{(m)}\sim \mathcal{N}(z|E_{L-1}), \hspace{3mm} E_{L-1}\supset E_L\equiv E,
\end{equation}
where $\mathcal{N}(z|A)\propto \mathcal{N}(z)I_A(z)$ denotes the standard multivariate normal distribution conditioned on $A$. By construction (this is the SubSim stopping criterion), at least $\tilde{m}=\beta m$ of these samples are in $E$. Let
\begin{equation}\label{eq:SSe}
Z^{(1)},\ldots,Z^{(m^*)}\sim\mathcal{N}(Z|E), \hspace{5mm} \beta m\leq m^*<m,
\end{equation}  
denote those samples. The conditional expectation can then be estimated as follows:
\begin{equation}\label{eq:expectedSN}
\mathbb{E}[S_N|Z\in E]\approx \widehat{\mathbb{E}}^{\mathbb{Q}}_{SubSim}=\frac{1}{m^*}\sum_{i=1}^{m^*}S_N(Z^{(i)}),
\end{equation}
where $S_N(Z^{(i)})=S_N(Z_1^{(i)},\ldots,Z_N^{(i)})$ is the final value of the asset price obtained from (\ref{eq:asset}). The expression in (\ref{eq:expectedSN}) in essence gives the expected terminal price of the underlying asset under the risk--neutral measure as the average of all the generated asset price paths. Combining (\ref{eq:SSp}) and (\ref{eq:expectedSN}), we obtain the SubSim estimate of the option price:
\begin{equation}\label{eq:estimate}
P\approx \widehat{P}_{SubSim}=\hat{p}_E^{SubSim}(\widehat{\mathbb{E}}^{\mathbb{Q}}_{SubSim}-K).
\end{equation}
SubSim as described above, yields an estimator for the execution probability $p_E$ which scales like a power of the logarithm of $p_E$ \citep{AB01}:
\begin{equation}\label{eq:COVSS}
\delta\left(\hat{p}_E^{SubSim}\right)=\sqrt{\frac{(1+\gamma)(1-\beta)}{M\beta(|\ln\beta|)^d}|\ln p_E|^d}\propto |\ln p_E|^{d/2},
\end{equation}
where $\gamma$ is a constant that depends on the correlation of the Markov chain states and $2\leq d\leq3$. 
Comparing (\ref{eq:COVSS}) against the CV of a standard MCS method ~\citep{Liu01,RC04} 
\begin{equation}\label{eq:COVMC}
\delta\left(\hat{p}_E^{MC}\right)=\frac{\sqrt{\mathrm{Var}\left[\hat{p}_E^{MC}\right]}}{\mathbb{E}\left[\hat{p}_E^{MC}\right]}=\sqrt{\frac{1-p_E}{Mp_E}} \propto p^{-1/2}_E
\end{equation}
reveals a serious drawback of MCS: it is inefficient in estimating small probabilities of rare events. Indeed, as $p_E \rightarrow 0$, then $\delta\left(\hat{p}_E^{MC}\right)\approx 1/\sqrt{Mp_E}$. This means that the number of samples $M$ needed to
achieve an acceptable level of accuracy is inversely proportional
to $p_E$, and therefore very large, $M\propto 1/p_E\gg1$. 
Therefore, for rare events, where probabilities are small $p_E\ll1$, the CV of SubSim is significantly lower than that of MCS, $\delta\left(\hat{p}_E^{SubSim}\right)\ll \delta\left(\hat{p}_E^{MC}\right)$. This property guaranties that SubSim produces more accurate (on average) estimates of small probabilities of rare events. 

In case the asset price $S$ has high volatility, then discrete asset price trajectories $S_1,\ldots, S_N$ will have large variability and with large probability will either cross the barriers and expire or end up bellow the strike. This means that having a positive payoff will be a rare event. This suggests -- and we confirm this by simulation in Section~\ref{sec5} -- that SubSim should be substantially more efficient in estimating prices of barrier options on high volatility assets than MC-based methods.

\section{Complexity Theorem}

\noindent The complexity theorem relates the execution probability $p_E$ with the mean squared error (MSE) and the computational complexity/cost of the SubSim estimator $\hat{P}$ for the option price $P$ at $t=0$, by examining their limiting behavior. The theorem does not make any assumptions regarding the underlying SDE or the functional of the solution used. 
\begin{theorem}
	The SubSim estimator $\hat{P}$ for a functional of the solution $\hat{S}$ to a given SDE has 
	\begin{enumerate}[label=(\roman*)]
		\item a MSE bounded from above by $c_1 \delta^2|\log p_E|^{-k}$, 
		
		\item with computational cost which has an upper bound of $c_2 \delta^{-2}|\log p_E|^r$,
	\end{enumerate}
	where $c_1, c_2$ are constants, $\delta$ is the CV of $\hat{P}$, $p_E$ is the probability of positive payoff at maturity and $r$ a parameter dependent on the correlation between the intermediate execution probabilities.    
\end{theorem}
\textbf{Proof.}
Using result (\ref{eq:COVSS}) we have that the squared CV of the execution probability $p_E$  is equal to
\begin{equation} \label{eq:covbasic}
\delta^2 = \frac{(1+\gamma)(1-\beta)}{\beta|\log \beta|^r Lm} |\log p_E|^r,
\end{equation}
where $\gamma$ is a constant related to the correlation between the states of the Markov chains used for the sampling at different levels, $\beta$ is the level probability, $L$ is the total number of subsets and $m$ represents the number of samples per subset (the product $Lm$ approximates the total number of samples $M$ in \eqref{eq:COVSS}). By \eqref{eq:estimate} we see that the option price estimate given by SubSim is a function of the execution probability $p_E$, the number of MMA samples that lead to a non-zero payoff and the payoff at maturity $S_N(Z^{(i)}) - K$.  As a result, the CV of the SubSim estimator $\hat{P}$ for the option price P is equal to the CV of $p_E$ times a scaling factor (the payoff at $t=T$) and the CV in \eqref{eq:covbasic} can be used. Now, the complexity of $\hat{P}$ given by the product of the samples per level times the number of simulation levels used is equal to
\begin{equation}
C = Lm = \frac{(1+\gamma)(1-\beta)}{\beta|\log \beta|^r \delta^2} |\log p_E|^r = \frac{(1+\gamma)(1-\beta)}{\beta \delta^2} |L|^r,
\end{equation}
by noting that the number of simulation levels $L$ is chosen as $L = \log p_E/\log \beta$. Fixing $\beta$ and treating $\gamma$ as a known constant we have that 
\begin{equation} \label{eq:compcost}
C \propto |L|^r \delta^{-2} \leq c_2 |L|^r \delta^{-2} \: \: \: \: or \: \: \: \: C \leq c_2 \delta^{-2}|\log p_E|^r,
\end{equation} 
which yields the upper bound of the computational complexity, given that $L$ is $O(|\log p_E|^r)$ for fixed $\beta$. Moreover, considering the definition for the coefficient of variation for $\hat{P}$ we have
\begin{equation}{\label{eq:covmse}}
\delta_{\hat{P}} = \frac{\sqrt{VAR[\hat{P}]}}{\mathbb{E}[\hat{P}]}= \frac{\sqrt{MSE[\hat{P}] - BIAS[\hat{P}, P]^2}}{\mathbb{E}[\hat{P}]}.
\end{equation}
Squaring both sides of (\ref{eq:covmse}) gives 
\begin{equation}
{\label{eq:covsq}}
\delta_{\hat{P}}^2 = \frac{MSE[\hat{P}] - BIAS[\hat{P}, {P}]^2}{\mathbb{E}[\hat{P}]^2},
\end{equation}
which equivalently can be written as 
\begin{equation}
MSE[\hat{P}] = \delta_{\hat{P}}^2 \mathbb{E}[\hat{P}]^2 + BIAS[\hat{P}, P]^2.
\end{equation}
Now, we use Propositions 1 and 2 \citep{AB01} which prove that both the bias and the squared CV $\delta^2$ of $p_E$ are bounded above by $c_3/m$. As a result, the first term of the $MSE$ is $O(1/m)$ while the second term is $O(1/m^2)$ which gives an $MSE$ bounded above by $1/m$ as for large values of $m$ it dominates the $O(1/m^2)$ term. 

By (\ref{eq:covbasic}) we also notice that $\delta^2$ is $O(|\log p_E|^r L^{-1} m^{-1})$ from which we obtain $m = O(|\log p_E|^r L^{-1} \delta^{-2})$. Setting $L = \log p_E / \log \beta = O(|\log p_E|)$ and fixing $\delta^2$, the number of samples $m$ becomes $O(|\log p_E|^{k})$ where $k = r-1 \leq 3$ is a new constant. Consequently, we end up with an MSE bounded from above by 
\begin{equation} \label{eq:msefin} 
MSE \triangleq \mathbb{E}[(P - \hat{P})^2] \leq c_1 \frac{1}{|\log p_E|^k}. \hspace{1em} \square
\end{equation} 

\begin{figure}[t]
	\caption{\textbf{Mean squared error and complexity/cost of $\hat{P}$.} The simulation results show that the MSE scales like $1/|\log p_E|^k$, where $ k=3 $, is a constant (left). In accordance with the theoretical findings, simulated MSE drops with increasing $p_E$. Computational cost/complexity of SubSim with respect to the probability of execution (right). The simulation results show that the cost can be bounded above by a function of $|\log p_E |^r$, $r =4$. The theoretical lines are calculated using the results in \cref{eq:compcost} and \cref{eq:msefin} with the CVs and execution probabilities of \cref{tab1}.}
	
	\centerline{\includegraphics[scale = 0.71]{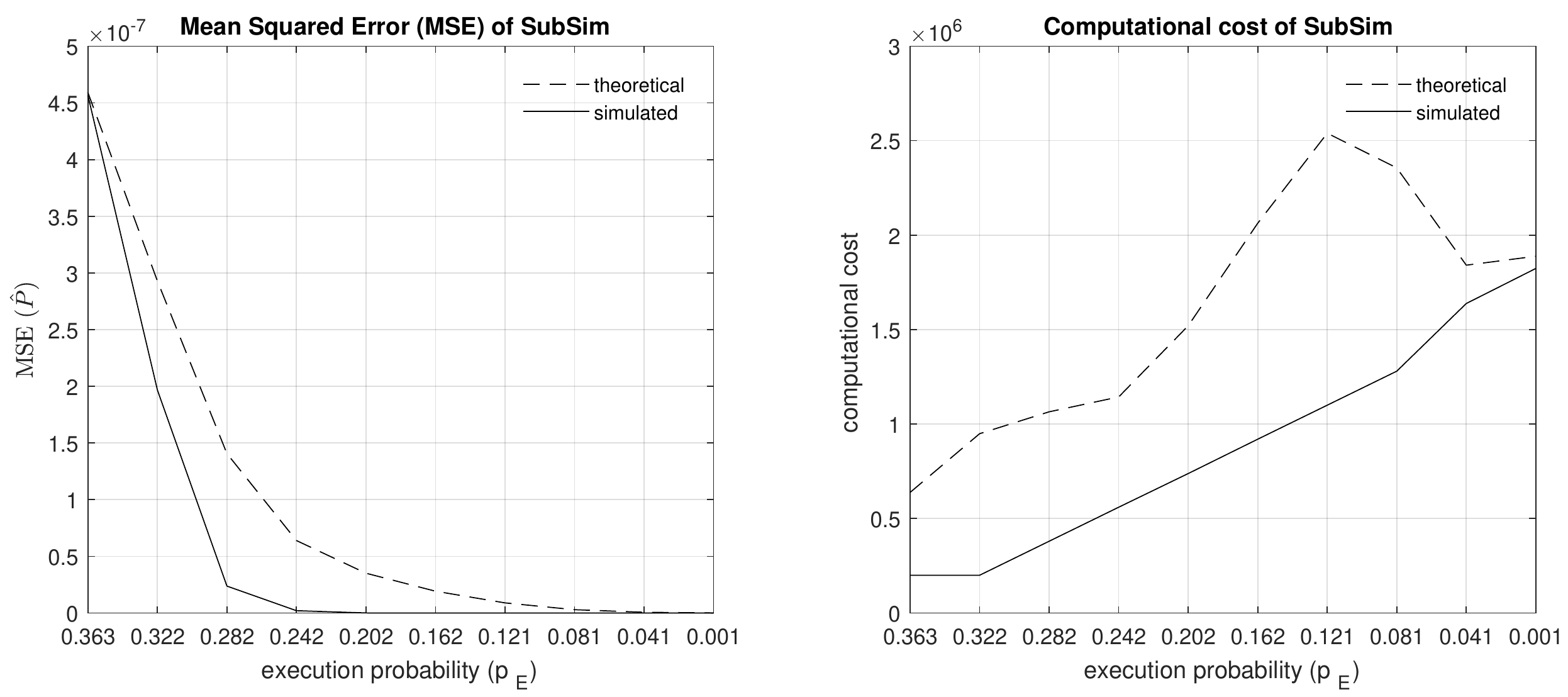}}
	
	\label{fig:mseandcost}
\end{figure}

The result in $(i)$ is very important as it shows that by decreasing the probability of contract execution (i.e., generating a more rare event) results in a smaller MSE while at the same time, the corresponding CV grows (see also results in Table \ref{tab1}). Moreover, in $(ii)$ we show that the computational complexity of SubSim is inversely proportional to the square of the target CV $\delta$ and the natural logarithm of the execution probability $p_E$. On one hand, as the target CV becomes smaller (i.e., we demand a more accurate output), the cost increases as the method uses more subsets and subsequently a larger number of samples. On the other hand, as the execution probability decreases, the absolute value of its logarithm increases, resulting in a higher computational cost as the lower the execution probability the more demanding the estimation of $\hat{P}$ becomes. Figure \ref{fig:mseandcost} shows the results of a simulation run (repeated 100 times) to compare how the MSE and the computational complexity scale with respect to $p_E$ according to the SubSim theory and the experimental outputs.  

\section{Simulation Study}\label{sec5}

\subsection{Barrier Options}
\noindent Our numerical experiments focus on pricing double knock-out barrier call options, but it is straightforward to extend the proposed methodology to other types of barrier options. Suppose that barriers are monitored during time period $[0,T]$ at equally spaced times $0=t_0<t_1<\ldots<t_N=T$ with frequency $\Delta t=T/N$,  and the option expires if the asset $S_t$ hits either the upper $U$ or the lower $L$ barrier. Let us denote the corresponding asset prices by $S_n=S_{t_n}$, the drift by $\mu_n=\mu(t_n)$ and the volatility by $\sigma_n=\sigma(t_n)$. 

\begin{figure}[t]
	\caption{\textbf{Asset price trajectories}. The top panel shows two asset trajectories that lead to a zero payoff: one trajectory breaks the upper barrier $U$ at time $t_k$, the other ends up below the strike, $S_N<K$. The bottom panel shows an asset price trajectory that results in a positive payoff $S_N-K$. For the sake of illustration, both lower and upper barriers are constant.}
	\centerline{\includegraphics[width=0.8\linewidth]{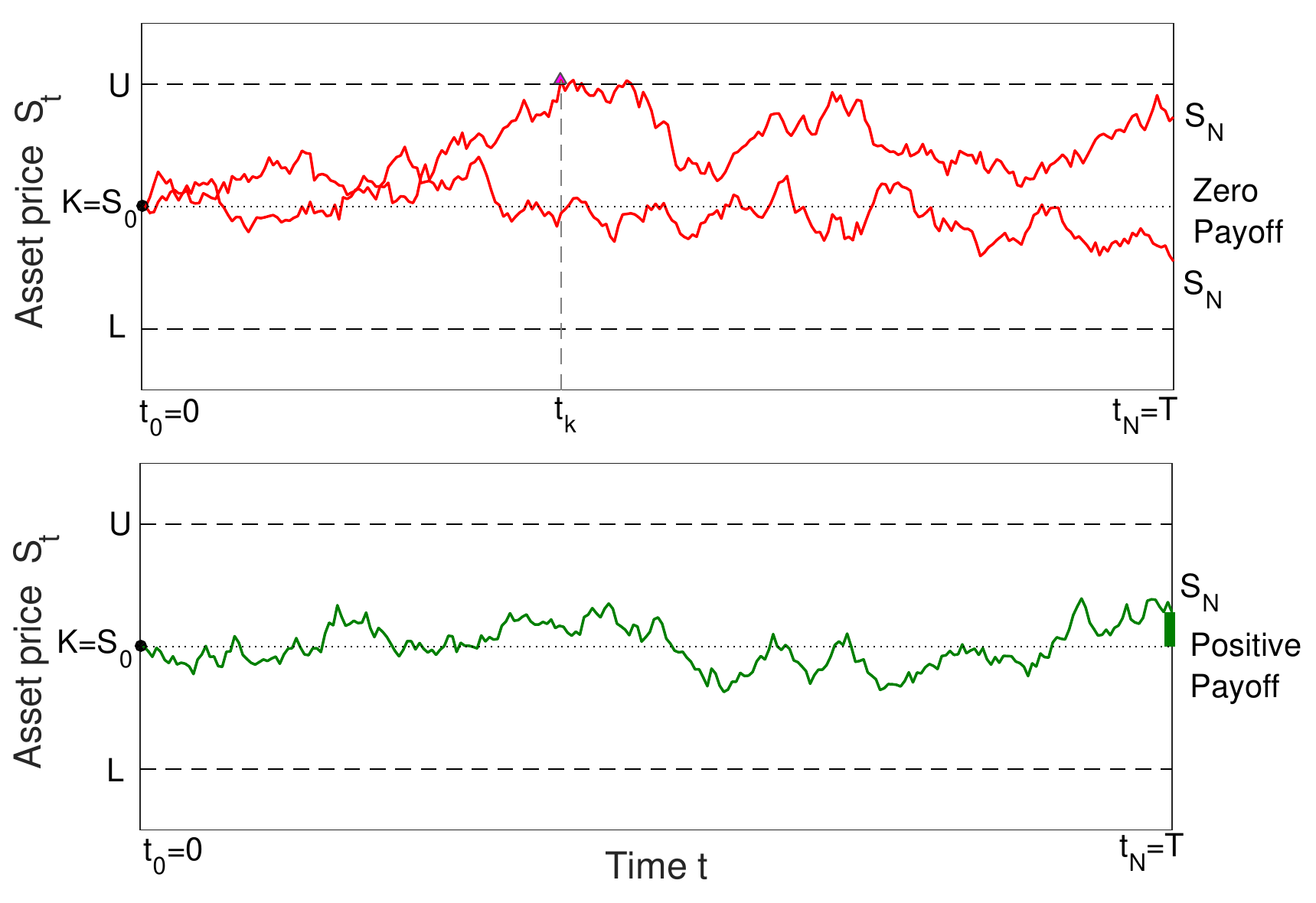}}
	
	\label{fig1}
\end{figure}

The quantity of interest is the barrier option price at the beginning of the contract ($t_0=0$), given by (\ref{eq:price}), which takes a non--zero value only in case the asset price trajectory remains within the two barriers. For illustrative purposes, Figure~\ref{fig1} shows several asset trajectories that lead to both option expiration and positive payoff.  

\subsection{Simulation results for SubSim vs standard MCS}\label{sec:SSvsMC}

In the first of our numerical experiments, we consider a double knock-out barrier call option with a starting price (spot) $S_0=100$, strike $K=100$, and constant lower and upper barriers $L=90$ and $U=110$. A double knock--out option expires worthless in case either the upper or the lower barrier is crossed by the asset price trajectory over the life of the option ($[0,T]$). In any other case, the payoff at maturity is calculated as a plain vanilla European call option (i.e., $P = (S_T - K)^+$, where $S_T$ is the terminal asset price). The option is discretely monitored during  time period $[0,T]$ at equally spaced times $0=t_0<t_1<\ldots<t_N=1$ with frequency $\Delta t=T/N$, where $N=250$ (approximate number of trading days in a financial year). We further assume that the drift of the underlying asset is constant $\mu=0.1$. To observe the effect of high volatility, we vary the value of $\sigma$ over ten different values logarithmically spaced between $\sigma_{\mathrm{min}}=0.2$ and  $\sigma_{\mathrm{max}}=0.4$. 

The quantity of interest, the fair option price at the beginning of the contract ($t_0=0$)  is given by
\begin{equation}
P_0=P\exp\left(-\int_0^Tr(t)dt\right),
\end{equation}  
where $P$ is the value of the option at the end of time period given by (\ref{eq:price}) and estimated by (\ref{eq:estimate}), $e^{-\int_0^Tr(t)dt}$ is the discounting factor from maturity $t_N = T$ to $t_0 = 0$, and $r(t)$ is the interest rate, which is assumed to be constant in this example, $r=0.1$.

\begin{figure}
	\caption{\textbf{Ratios of CVs.} The ratios $\delta(\hat{p}_E^{MCS})/\delta(\hat{p}_E^{SubSim})$ and $\delta(\widehat{P}_0^{MCS})/\delta(\widehat{P}_0^{SubSim})$ versus the volatility $\sigma$ are presented (left). Total number of samples used in Subset Simulation when $L = 90$ and $U = 110$ (right).}
	\centerline{\includegraphics[scale = 0.74]{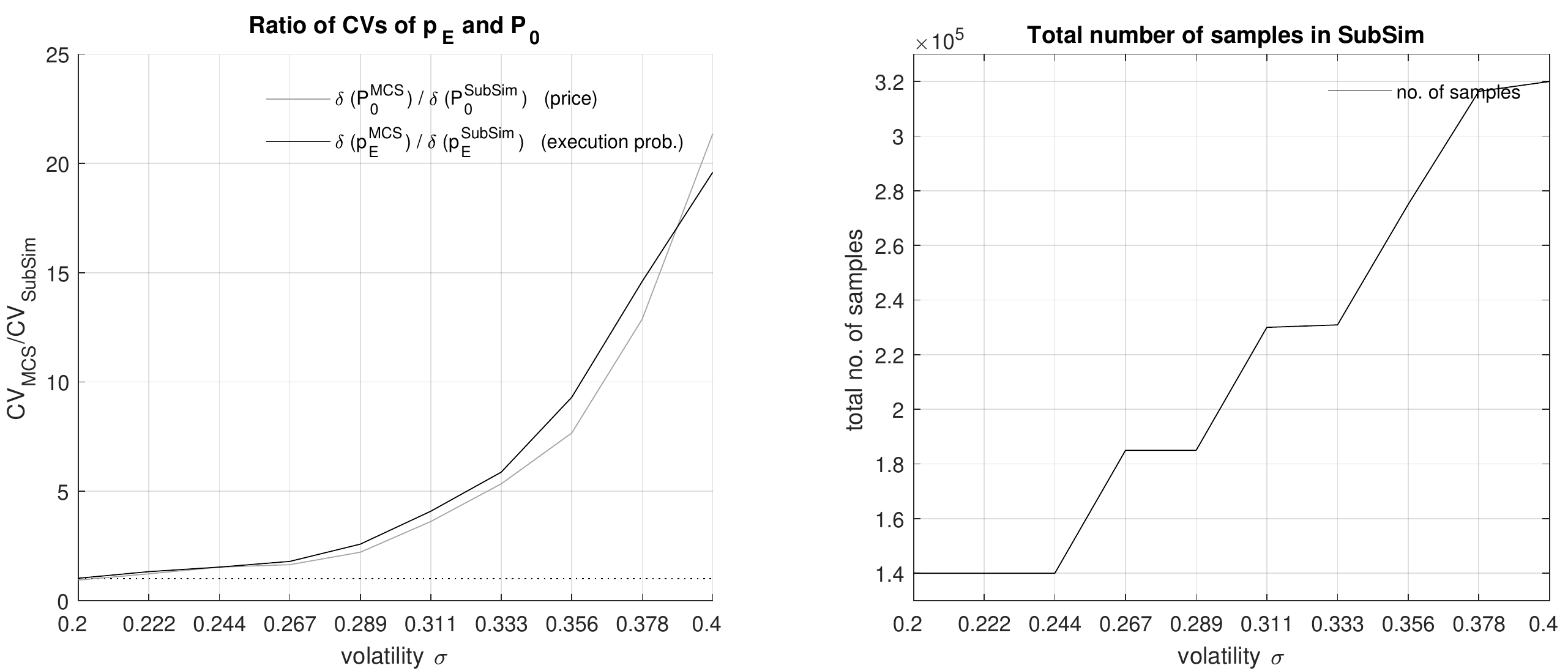}}
	\label{fig6}
\end{figure}

First, we use SubSim with $m=50,000$ samples per subset to estimate both the probability $p_E$ of having a positive payoff at the end of the period, $p_E\approx \hat{p}_E^{SubSim}$, and the option price,
\begin{equation}
P_0\approx \widehat{P}_0^{SubSim}=\widehat{P}_{SubSim}e^{-rT}. 
\end{equation}
The mean values of estimates and their CVs computed from 100 independent runs of the SubSim algorithm are presented in Table~\ref{tab1}. As expected, as the asset volatility $\sigma$ increases, the event of having a positive payoff becomes increasingly rare (e.g. if $\sigma=0.4$, then $p_E\approx\num{2E-07}$) and, as a result, the option becomes cheaper. The right plot in Figure~\ref{fig6} shows the average (based on 100 runs) total number of samples $M$ used by SubSim versus the volatility $\sigma$. The obtained trend is again expected: as $\sigma$ increases, the probability $p_E$ becomes smaller, and, therefore, the number $L$ of subsets in (\ref{eq:SSestimate}) increases, which leads to the increase in the total number of samples (\ref{eq:totalSS}).  

\begin{table}
	\footnotesize
	\caption{\textbf{Simulation results}. This table shows the mean values and coefficients of variations of the estimates of the execution probability $p_E$ and the barrier option price $P_0$, obtained by SubSim and MCS for different values of volatility $\sigma$. All statistics are obtained from 100 independent runs of the algorithms.}
	\begin{center}
		{\begin{tabular}{ccccc}
				\hline
				$\sigma$  & $\hat{p}_E^{SubSim} / \hat{p}_E^{MCS}$ & $\widehat{P}_0^{SubSim} / \widehat{P}_0^{MCS}$ & $\delta(\hat{p}_E^{SubSim}) / \delta(\hat{p}_E^{MCS}) $ & $\delta(\widehat{P}_0^{SubSim}) / \delta(\widehat{P}_0^{MCS})$\\ 
				\hline 
				0.200  &  \num{8.30E-03} / \num{8.26e-03} & \num{2.93E-02}  / \num{2.91E-02}  & 0.030 / 0.0281  & 0.034 / 0.0347 \\ 
				0.216  &  \num{4.32E-03} / \num{4.34E-03} &  \num{1.52E-02} / \num{1.53E-02}  & 0.032 / 0.0391  &  0.036 / 0.0476 \\  
				0.233  &  \num{2.04E-03} / \num{2.04E-03} &  \num{7.18E-03} / \num{7.19E-03}  &  0.039 / 0.0596 &  0.044 / 0.0673 \\ 
				0.252  &  \num{8.67E-04} / \num{8.76E-04} &  \num{3.06E-03} / \num{3.11E-03}  &  0.048 / 0.0788 &  0.055 / 0.0985 \\ 
				0.272  &  \num{3.23E-04} / \num{3.21E-04} &  \num{1.14E-03} / \num{1.15E-03}  &  0.057 / 0.126  &  0.062 / 0.160 \\ 
				0.294  &  \num{1.06E-04} / \num{1.08E-04} &  \num{3.75E-04} / \num{3.80E-04}  &  0.060 / 0.217  &  0.069 / 0.282 \\ 
				0.317  &  \num{2.91E-05} / \num{2.63E-05} &  \num{1.03E-04} / \num{9.38E-05}  &  0.076 / 0.406  & 0.081 / 0.476 \\ 
				0.343  &  \num{6.85E-06} / \num{5.66E-06} &  \num{2.46E-05} / \num{2.14E-05}  & 0.099 / 0.759   & 0.109 / 1.014 \\ 
				0.370  &  \num{1.31E-06} / \num{9.93E-07} &  \num{4.69E-06} / \num{3.06E-06}  &  0.153 / 1.971  &  0.160 / 2.337 \\
				0.400  &  \num{1.99E-07} / \num{2.45E-07} &  \num{7.20E-07} / \num{1.10E-06}  &  0.180 / 3.844  & 0.205 / 4.017\\ 
				\hline
		\end{tabular}}
		\label{tab1}
	\end{center}
\end{table}

Next, we use MCS to estimate $p_E$ and $P_0$. To ensure fair comparison of the two methods, for each value of $\sigma$, MCS is implemented with the same total number of samples as in SubSim.  The mean values of Monte Carlo estimates for the execution probability $\hat{p}_E^{MCS}$ and the option price $\widehat{P}_0^{MCS}=\widehat{P}^{MCS}e^{-rT}$, with their CVs are presented in Table~\ref{tab1}. The mean values of $\hat{p}_E^{MCS}$ and $\widehat{P}_0^{MCS}$ are approximately the same as those of $\hat{p}_E^{SubSim}$ and $\widehat{P}_0^{SubSim}$, which confirms that SubSim estimates are approximately unbiased. The CVs, however, differ drastically. Namely, $\delta(\hat{p}_E^{SubSim})$  and $\delta(\widehat{P}_0^{SubSim})$ are  substantially smaller than $\delta(\hat{p}_E^{MCS})$ and  $\delta(\widehat{P}_0^{MCS})$, respectively. This effect is more pronounced the larger the volatility. For example, if $\sigma=0.4$, then SubSim is approximately 20 times more efficient than MCS, i.e., on average, SubSim produces 20 times more accurate estimates, where the accuracy is  measured by the CV. As explained at the end of Section~\ref{sec3}, this result stems from the fact that SubSim is more efficient than MCS in estimating small probabilities of rare events, and if volatility is large, then the event of having a positive payoff is rare.

To visualize how SubSim outperforms MCS as the volatility increases, in the left plot of Figure~\ref{fig6} we plot the ratios of CVs $\delta(\hat{p}_E^{MCS})/\delta(\hat{p}_E^{SubSim})$ and $\delta(\widehat{P}_0^{MCS})/\delta(\widehat{P}_0^{SubSim})$ versus $\sigma$.  Since the mean values of SubSim and MCS estimates are approximately the same, the ratios of CVs are approximately the ratios of the corresponding standard errors. Graphically, the cases where SubSim outperforms MCS for the estimation of the execution probability and the option price are those for which the corresponding value of $\delta(\hat{p}_E^{MCS})/\delta(\hat{p}_E^{SubSim})$ or $\delta(\widehat{P}_0^{MCS})/\delta(\widehat{P}_0^{SubSim})$ lies above the horizontal line $y=1$ (dotted line in Figure \ref{fig6}). At that level, both methods would exhibit the same level of accuracy measured by the CV, since $\delta_{MCS}$ would equal $\delta_{SubSim}$. We notice that SubSim outperforms MCS in every examined case as both lines (for $\hat{P_0}$ and $\hat{p_E}$) lie above the $y=1$ level.   

\begin{figure}
	\centering
	\caption{\textbf{Ratios of CVs of the option price $P_0$.} The results are plotted with respect to asset volatility, for Subset Simulation against Monte Carlo (top) and Subset Simulation against multilevel Monte Carlo (bottom). Four different barrier levels are presented (to perform the simulations we use mainly the codes provided by Mike Giles at https://people.maths.ox.ac.uk/gilesm/mlmc/ doing the necessary adjustments in file mcqmc06.m).}
	\includegraphics[scale = 0.97]{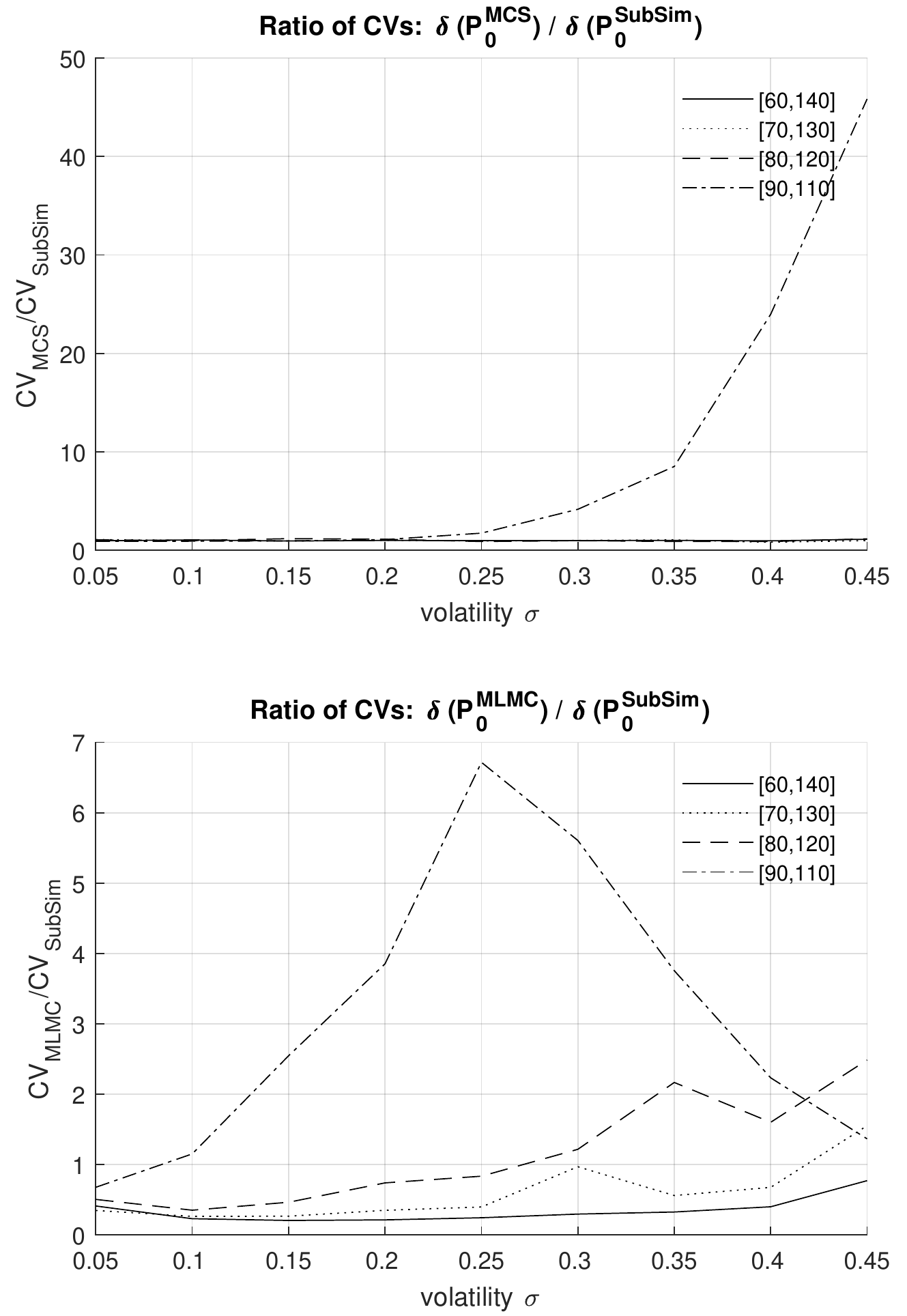}
	
	\label{fig7}
\end{figure}

In the second of our simulation tests we increase the number of samples to $m = 200,000$ using also different levels for the lower and the upper barrier.  The reason we consider more samples is to compare SubSim against not only MCS but also multilevel Monte--Carlo (see subsection \ref{sec:SSvsMLMC}), where $m = 200,000$ is considered in the original barrier option numerical experiments. To maintain a fair comparison we perform our MCS tests with the same number of samples as in SubSim.  The top graph of Figure~\ref{fig7} plots the ratio of CV between SubSim and standard MCS with respect to the volatility of the underlying asset for four levels of the upper and lower barrier. It is immediately noticeable that for volatility values up to $0.25$ the two methods have comparable CVs (SubSim outperforms standard MCS as reported in Table \ref{t:COV1} but not significantly), providing evidence that for low--volatility assets the two methods produce sufficiently accurate results. This result is not surprising as SubSim is designed by construction to deal with problems with extremely small execution probabilities.  

However, as volatility increases, SubSim outperforms naive MCS in all barrier levels, while especially in the case of $L=90$ and $U=110$ (barriers close to $S_0$) and $\sigma \geq 0.40$ (a high--volatility asset), SubSim is up to 50 times more efficient than standard MC; for lower levels of $\sigma$, SubSim still outperforms MCS. 

\subsection {Simulation results for SubSim vs MLMC} \label{sec:SSvsMLMC}

In this section we compare the performance of SubSim against the multilevel Monte Carlo method \citep{giles2008multilevel, giles2008improved}, when both used to price a double knock--out barrier call option with two fixed barriers set at four different levels, while all the other parameters remain the same as in subsection \ref{sec:SSvsMC}. The original multilevel MCS method was developed to price single knock--out barrier options, amongst other exotic derivatives, and thus we add a component for the second barrier in order to accommodate double barrier options as well (see appendices \ref{appendixA} and \ref{appendixB}). 

The price at $t=0$ of the asset is $S_0 = 100$, the strike price is $K = 100$ and the time--increment is $\Delta t = h = T/n$ where $n$ represents the number of discrete monitoring points of the barrier option. In the case of MLMC, $n$ varies between levels as it is a function of a constant $M$ and level $l$, where $l = 0,1,2,\dots,L$. The barriers take four different values in increments of ten between $60$ and $90$ (lower) and $110$ and $140$ (upper). The drift of the diffusion equation is equal to $\mu = 0.10$, while the volatility (diffusion coefficient) varies between $0.05$ and $0.45$ taking nine discrete values linearly spaced in this interval. Finally, the  risk--free rate at which we discount the terminal payoffs is known and fixed at $r = 0.10$. 

The bottom graph of Figure~\ref{fig7} plots the ratio of CV between SubSim and MLMC for four levels of barriers against asset's volatility. For barriers which lie far from the price of the asset at $t = 0$ (i.e., $[60,140]$ and $[70,130]$ represented by the solid and the dotted line respectively), MLMC produces more accurate results than SubSim. Nevertheless, we notice that as asset volatility increases the performance of SubSim improves, approaching that of MLMC without surpassing it. SubSim outperforms MLMC when $L = 90$ and $U=110$ (dashed/dotted line) and when $L = 80$ and $U=120$ (dashed line) and the volatility of the underlying asset is higher than 0.25. In both cases, the probability of a non--zero payoff at $t=T$ is extremely small (Table \ref{tab1}), and hence the use of SubSim provides more accurate results compared either to standard MCS or MLMC. The evidence we obtain here further supports the findings in Section \ref{sec:SSvsMC} that SubSim is an efficient technique to price barrier options on high volatility assets, especially when the barriers are close to the initial price of the underlying asset.  

Exact values for $\hat{P}_0^{\{MCS,MLMC,SubSim\}}$ (option price at $t=0$ for each of the three methods) and $CV_{P_0}^{\{MCS,MLMC,SubSim\}}$ can be found in Tables \ref{t:optprice} and \ref{t:COV1}, respectively in appendix \ref{appendixC}. For visualization purposes, we also plot these results in Figures \ref{fig:optval} and \ref{fig:cov}.
		
\section{Conclusion}\label{sec6}

In this paper, we develop a new stochastic simulation method for pricing barrier options. The method is based on Subset Simulation (SubSim), a very efficient algorithm for estimating small probabilities of rare events. The key observation allowing to exploit the efficiency of SubSim is that the barrier option price can be written as a function of the probability of option execution and a certain conditional expectation, which can both be estimated efficiently by SubSim. In the case of barrier options on high--volatility assets, SubSim is especially advantageous because of the very small probability of the contract to remain valid until maturity. We first compare the proposed SubSim method against the standard Monte Carlo simulation (MCS) to show that SubSim always outperforms MCS, confirming this with a  series of numerical examples. Moreover, we show that the higher the volatility of the underlying asset (i.e. the smaller the probability of option execution), the larger the advantage of SubSim over MCS. Next, we compare our proposed method with the multilevel Monte--Carlo (MLMC) simulation introduced in \cite{giles2008multilevel}. Although MLMC outperforms SubSim in general, we find that SubSim can still be more efficient than MLMC, -- where efficiency is measured by the coefficient of variation (CV) -- in cases where the volatility of the underlying asset is high and the barriers are set close to the starting price of the asset. As a result, the method we propose here complements MLMC,  handling special cases of barrier option settings more efficiently.   
	



\newpage
\bibliographystyle{plainnat} 
\bibliography{bib} 

\newpage
\appendix
\appendixpage
\footnotesize
\section{MMA sampling from the target distribution $f_z$} \label{appendix0}
To sample from the target distribution $f_z(z|E_{i-1})$, the MMA generates a Markov chain with stationary distribution $f_z(z|E_{i-1})$. Namely, if we let $Z^{(j)}\in E_{i-1}$ be the current state, then the next state $Z^{(j+1)}$ is generated as follows:
\begin{enumerate}
	\item Generate a candidate state $\Upsilon=(\Upsilon_1,\ldots,\Upsilon_N)$:
	\begin{enumerate}
		\item For each $k=1,\dots,N$, generate $\Psi_k\sim q(\psi|U^{(j)}_k)$, where $q$ is a symmetric, $q(\psi|u)=q(u|\psi)$, univariate proposal distribution, e.g. Gaussian distribution centered at $U^{(j)}_k$, the $k^{\mathrm{th}}$ component of $U^{(j)}$. 
		\item Compute the acceptance probability: 
		\begin{equation}
		a_k=\min\left\{1, \frac{f_k(\Psi_k)}{f_k(U^{(j)}_k)}\right\},
		\end{equation} 
		where $f_k$ is the marginal PDF of $U_k$, $f_U(u)=\prod_{k=1}^{N}f_k(u_k)$, and $U_1,\ldots, U_N$ are assumed to be independent. 
		\item Set 
		\begin{equation}
		\Upsilon_k=\begin{cases}
		\Psi_k, & \mbox{with probability } a_k,\\
		U^{(j)}_k & \mbox{with probability } 1-a_k.
		\end{cases}
		\end{equation}
	\end{enumerate}
	\item Accept or reject the candidate state:
	\begin{equation}
	U^{(j+1)}=
	\begin{cases}
	\Upsilon, & \mbox{if } \Upsilon\in E_{i-1},\\
	U^{(j)}, & \mbox{if } \Upsilon\notin E_{i-1}.
	\end{cases}
	\end{equation}
\end{enumerate}

\section{Probability of survival of a barrier option} \label{appendixA}
\noindent The pricing of barrier options is a first passage time problem in which we are interested in the first time that the price trajectory of the underlying asset crosses a prespecified barrier. Now, assuming that $U>S_0$ and $L<S_0$ are the upper and lower barriers respectively, the survival indicator function of the barrier option in (\ref{eq:price}) can be approximated via its discrete form
\begin{equation}\label{eq:survind}
\prod_{i=0}^{n-1} \mathbf{\textit{I}}_{\{\hat{M}_i \leq U \: \land \: \hat{m}_i \geq L\}}
\end{equation}
where $\hat{M}_i$ and $\hat{m}_i$ are the maximum and minimum, respectively, of (\ref{eq:asset}) in $[0,nh]$ and $T=nh$ or $h=T/n$ is the size of the timestep on a discrete grid. Equation (\ref{eq:survind}) takes the value one if and only if the conditions for $\hat{M}_i$ and $\hat{m}_i$ are met at every time--step of the discretized problem, otherwise the product ~(\ref{eq:survind}) becomes zero and the option expires worthless. Following \cite{glasserman2013monte} (see particularly section 6.4 and example 2.2.3) we sample the minimum and the maximum of $S$ by formulating the following problem:  
\begin{equation}\label{eq:maxX}
M(t) = \max_{0 \leq u \leq t}{S(u)}
\end{equation} 
with 
\begin{equation} \label{eq:maxEuler}
\hat{M}^h(n) = \max\{S(0), S(h), S(2h),\dots, S(nh) \}
\end{equation}
the maximum of the approximation of S on $[0, nh]$, and
\begin{equation}
m(t) = \min_{0 \leq u \leq t}{S(u)}
\end{equation}
with 
\begin{equation} \label{eq:maxEuler2}
\hat{m}^h(n) = \min\{S(0), S(h), S(2h),\dots, S(nh) \}
\end{equation}
the minimum of a discrete time approximation of S on $[0, nh]$.

In the sampling of the maximum, conditioning on the endpoints $S(0)$ and $S(T)$, the process $\{S(t), 0 \leq t \leq T \}$ becomes a Brownian bridge, and thus we sample from the distribution of the maximum of a Brownian bridge, a Rayleigh distribution, which results in 
\begin{equation}
M(T) = \frac{S(T) + \sqrt{S(T)^2 - 2T\log X}}{2},
\end{equation} 
where $X$ is a uniformly distributed random variable in $[0,1]$. Now, let $\hat{S}_{ih}$ be a discrete time approximation of the solution of $S$ in (\ref{eq:brown}), where $i = 0,1, \dots, n$, $h = T/n$. To obtain a good estimation for $\hat{M}^h$ (i.e. the maximum of the interpolating Brownian bridge) and decrease the error induced by the discretization (i.e., the case where ${S}_u$ crosses $U$ or $L$ between two grid points), we interpolate over $[ih, (i+1)h]$, which given the end points $S_{i}$ and $S_{i+1}$ results in 
\begin{equation}\label{eq:bbmax}
M_i = \frac{S(i) + S(i+1) + \sqrt{[S(i+1) - S(i)]^2 - 2b^2h \log X}}{2}
\end{equation}  
with $X \sim \text{Unif} [0,1]$. 

\noindent Given a barrier $U$, the probability of survival for the option (the maximum price of the underlying asset to remain below $U$) in the fine--path estimation is given by 
\begin{equation}
\hat{p}_{i,U} = P(\hat{M}_i \leq U | \hat{S}_i, \hat{S}_{i+1}) = 1 - \exp \bigg(- \frac{2(U - \hat{S}_{i})(U - \hat{S}_{i+1})}{b^2h}  \bigg), 
\end{equation} 
where $b$ is the fixed standard deviation of the underlying asset price and $h$ is the time--step in the discretization process. The corresponding estimation for a coarse--path  is equal to 
\begin{multline} \label{eq:coarse1}
\hat{p}_{i,U} = P(\hat{M}_i \leq U | \hat{S}_i, \hat{S}_{i+1}) =\bigg\{ 1 - \exp \bigg(- \frac{2(U - \hat{S}_{i})(U - \hat{S}_{i+1/2})}{b^2h}      \bigg) \bigg\} \\ \times \bigg\{ 1 - \exp \bigg(- \frac{2(U - \hat{S}_{i+1/2})(U - \hat{S}_{i+1})}{b^2h}      \bigg) \bigg\}. 
\end{multline} 

\newpage
\section{Minimum of Brownian bridge} \label{appendixB}

\noindent We now derive analytically the probability of survival for a double barrier option in a fine path estimation, by calculating also the probability of the minimum of $\hat{S}$ to cross the lower barrier $L$. Conditioning on endpoints $\hat{S}_i$ and $\hat{S}_{i+1}$, the distribution of the minimum of the Brownian bridge (interpolated over $[i, (i+1)h]$) is given by
\begin{equation}\label{eq:bbmin}
m_i = \frac{S(i) + S(i+1) - \sqrt{[S(i+1) - S(i)]^2 - 2b^2h \log X}}{2},
\end{equation} 
where  $X \sim \text{Unif} [0,1]$.
Subsequently, the probability of the minimum $m_i$ of $\hat{S}$ to cross the lower barrier $L$ is equal to 
\begin{multline}\label{eq:minBB}
P(\hat{m}_i \leq L | \hat{S}_i, \hat{S}_{i+1} )\\ = P \bigg(\frac{\hat{S}(i) + \hat{S}(i+1) - \sqrt{[\hat{S}(i+1) - \hat{S}(i)]^2 - 2b^2h \log X}}{2} \leq L | \hat{S}_i, \hat{S}_{i+1}\bigg) \\
=   P \bigg(\sqrt{[\hat{S}(i+1) - \hat{S}(i)]^2 - 2b^2h \log X} \geq (\hat{S}(i) + \hat{S}(i+1)) - 2L | \hat{S}_i, \hat{S}_{i+1} \bigg) \\
=  P \bigg(\hat{S}(i+1)^2 - 2\hat{S}(i)\hat{S}(i+1) + \hat{S}(i)^2 - 2b^2h \log X \\\geq (\hat{S}(i) + \hat{S}(i+1))^2 - 4L(\hat{S}(i) + \hat{S}(i+1)) + 4L^2 | \hat{S}_i, \hat{S}_{i+1} \bigg) \\
=  P \bigg(\hat{S}(i+1)^2 - 2\hat{S}(i)\hat{S}(i+1) + \hat{S}(i)^2 - 2b^2h \log X \\\geq \hat{S}(i)^2 + \hat{S}(i+1)^2 + 2\hat{S}(i)\hat{S}(i+1) - 4L(\hat{S}(i) + \hat{S}(i+1)) + 4L^2 | \hat{S}_i, \hat{S}_{i+1} \bigg) \\
= P \bigg(-b^2h \log U \geq 2\hat{S}(i)\hat{S}(i+1) - 2L\hat{S}(i) + 2L\hat{S}(i+1) + 2L^2 | \hat{S}_i, \hat{S}_{i+1} \bigg) \\
= P \bigg(\log U \leq -\frac{2\hat{S}_i(\hat{S}_{i+1} - L) -2L(\hat{S}_{i+1} - L )}{b^2h} | \hat{S}_i, \hat{S}_{i+1} \bigg) \\ = P \bigg (\log U \leq -\frac{2(\hat{S}_i - L)(\hat{S}_{i+1} - L)}{b^2h} | \hat{S}_i, \hat{S}_{i+1} \bigg) \\
= P \bigg( U \leq \exp \bigg(-\frac{2(\hat{S}_i - L)(\hat{S}_{i+1} - L)}{b^2h} \bigg) | \hat{S}_i, \hat{S}_{i+1} \bigg) = \exp \bigg(-\frac{2(\hat{S}_i - L)(\hat{S}_{i+1} - L)}{b^2h} \bigg).
\end{multline}   
The probability in (\ref{eq:minBB}) refers to the case of the running minimum crossing the lower barrier. The probability to remain above the lower barrier is thus equal to its complement 
\begin{equation}
\hat{p}_{i,L} = 1 - \exp \bigg(-\frac{2(\hat{S}_i - L)(\hat{S}_{i+1} - L)}{b^2h} \bigg),
\end{equation}
and the probability of the asset price to remain within the barriers on $[0,T]$ is equal to 
\begin{equation} \label{eq:probsurv1}
\hat{p}_{i} = \hat{p}_{i,U}\hat{p}_{i,L} = \bigg\{1 - \exp \bigg(- \frac{2(U - \hat{S}_{i})(U - \hat{S}_{i+1})}{b^2h}      \bigg) \bigg\} \bigg\{1 - \exp \bigg(- \frac{2(\hat{S}_{i} - L)(\hat{S}_{i+1} - L)}{b^2h} \bigg)  \bigg\}. 
\end{equation}
The calculation of the probability of survival for the coarse path estimation follows trivially from (\ref{eq:probsurv1}) by adjusting it using (\ref{eq:coarse1}).
Then, the option remains alive until time $t = T = nh$ when  the asset price is bounded between $L$ and $U$, which in the case of a coarse path estimation, using a midpoint equal to $i+1/2$, equals
\begin{align}
\hat{p}_{i} &= \bigg\{ 1 - \exp \bigg(- \frac{2(U - \hat{S}_{i})(U - \hat{S}_{i+1/2})}{b^2h}      \bigg) \bigg\} \bigg\{ 1 - \exp \bigg(- \frac{2(U - \hat{S}_{i+1/2})(U - \hat{S}_{i+1})}{b^2h}      \bigg) \bigg\} \\ &\times \bigg\{ 1 - \exp \bigg(- \frac{2(\hat{S}_{i} - L)(\hat{S}_{i+1/2} - L)}{b^2h}  \bigg) \bigg\}  \bigg\{ 1 - \exp \bigg(- \frac{2(\hat{S}_{i+1/2} - L)(\hat{S}_{i+1} - L)}{b^2h}      \bigg) \bigg\}. 
\end{align}

\section{Simulation study results}\label{appendixC}
\begin{figure}[h]
	\caption{\textbf{Barrier option prices.} Results reported for the three methods with respect to volatility. The four graphs correspond to different levels of the upper and lower barrier.} \label{fig:optval}
	\centerline{\includegraphics[scale=0.85]{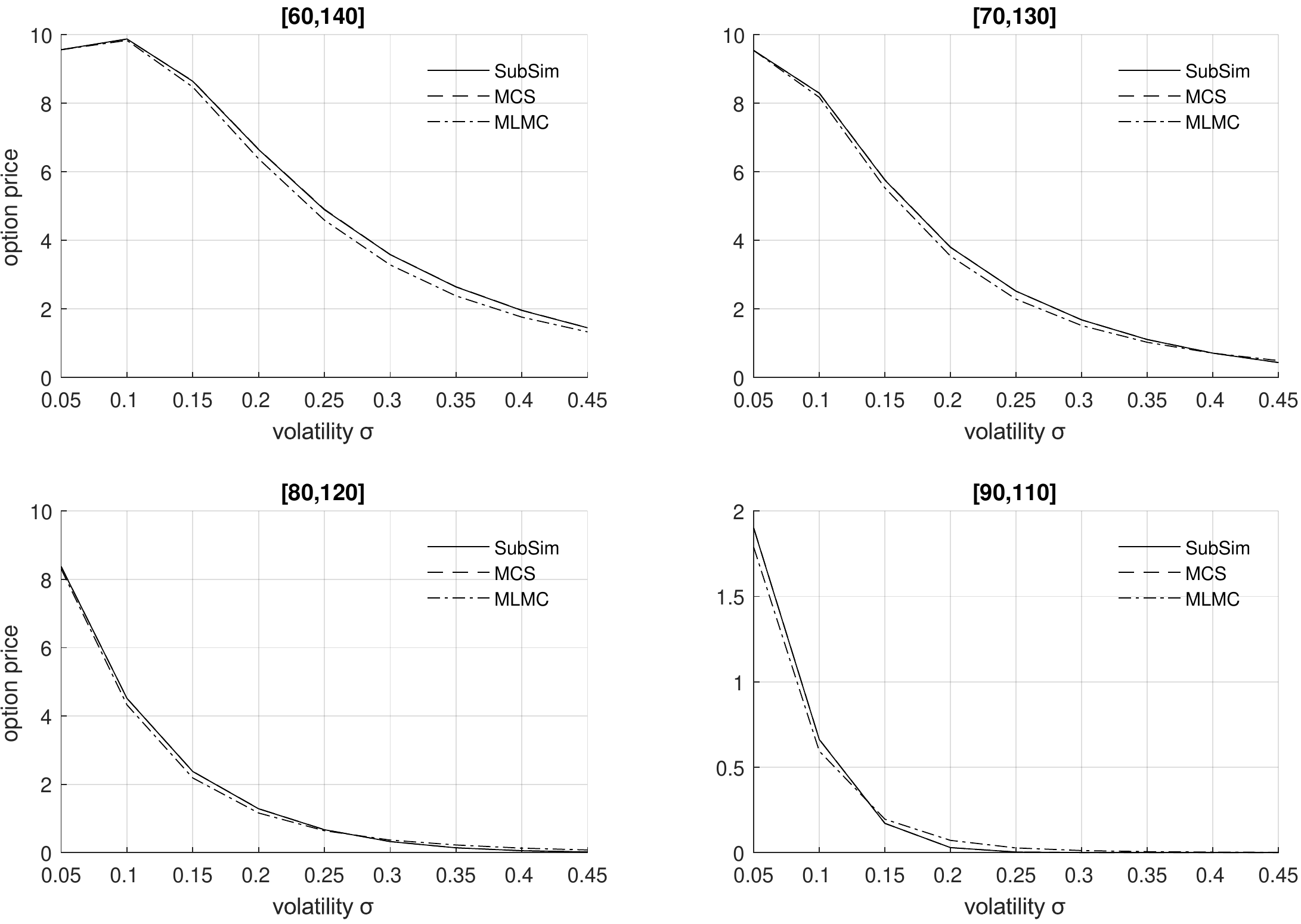}}
\end{figure}

\begin{figure}[h]
	\caption{\textbf{Coefficient of variation (CV).} Results reported for the three methods with respect to volatility for 100 runs of the pricing algorithm. The four graphs correspond to different levels of the upper and lower barrier.}
	\centerline{\includegraphics[scale=0.85]{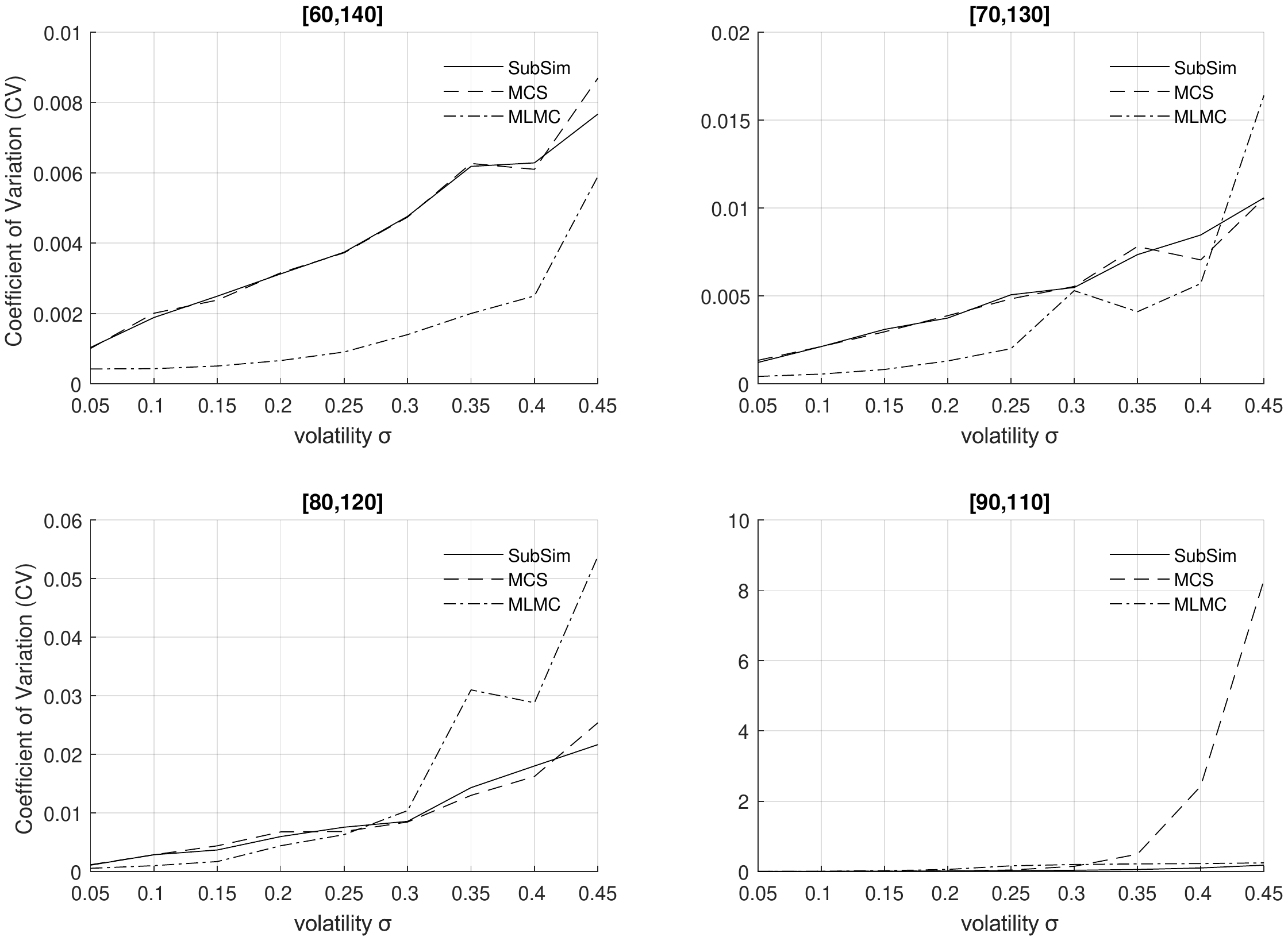}}
	\label{fig:cov}
\end{figure}

\pagestyle{plain}

\begin{table}[t]
	\centering	
	\caption{\textbf{Barrier option prices.} A comparison of the option prices derived by each of the three methods (MCS, MLMC and SubSim) for four barrier levels against volatility. }
	\begin{tabular}{llllll}
		\toprule
		& \multicolumn{5}{c}{Barriers}                                         \\
		&          & {[}60,140{]} & {[}70,130{]} & {[}80,120{]} & {[}90,110{]} \\
		\midrule
		Volatility ($\sigma$) & Method   &              &              &              &              \\
		\midrule
		0.05                  & Standard MCS & 9.5559       & 9.5345       & 8.3761       & 1.9009       \\
		& MLMC     & 9.5549       & 9.5339       & 8.3008       & 1.7882       \\
		& SubSim       & 9.5573       & 9.5351       & 8.3728       & 1.8997       \\
		0.10                  & Standard MCS & 9.8679       & 8.2903       & 4.5155       & 0.6617       \\
		& MLMC     & 9.8271       & 8.1682       & 4.3242       & 0.5941       \\
		& SubSim       & 9.8656       & 8.2862       & 4.5137       & 0.6615       \\
		0.15                  & Standard MCS & 8.6454       & 5.7592       & 2.3743       & 0.1712       \\
		& MLMC     & 8.4688       & 5.5283       & 2.1859       & 0.1956       \\
		& SubSim       & 8.6413       & 5.7570       & 2.3734       & 0.1711       \\
		0.20                  & Standard MCS & 6.6578       & 3.8014       & 1.2839       & 0.0290       \\
		& MLMC     & 6.3772       & 3.5392       & 1.1595       & 0.0716       \\
		& SubSim       & 6.6477       & 3.7958       & 1.2839       & 0.0292       \\
		0.25                  & Standard MCS & 4.8993       & 2.5194       & 0.6712       & 0.0033       \\
		& MLMC     & 4.5896       & 2.2841       & 0.6406       & 0.0273       \\
		& SubSim       & 4.8970       & 2.5148       & 0.6707       & 0.0033       \\
		0.30                  & Standard MCS & 3.5877       & 1.6833       & 0.3226       & 0.0003       \\
		& MLMC     & 3.2844       & 1.5152       & 0.3668       & 0.0120       \\
		& SubSim       & 3.5840       & 1.6792       & 0.3223       & 0.0003       \\
		0.35                  & Standard MCS & 2.6423       & 1.1106       & 0.1406       & 1.33E-05     \\
		& MLMC     & 2.3811       & 1.0275       & 0.2233       & 5.70E-03     \\
		& SubSim       & 2.6414       & 1.1096       & 0.1403       & 1.61E-05     \\
		0.40                  & Standard MCS & 1.9638       & 0.7114       & 0.0554       & 1.84E-06     \\
		& MLMC     & 1.7620       & 0.7101       & 0.1306       & 3.00E-03     \\
		& SubSim       & 1.9604       & 0.7107       & 0.0554       & 7.19E-07\\
		0.45       & Standard MCS & 1.4525	  &0.4387	  &0.0199	  &5.79E-08\\
		& MLMC        & 1.3312	&0.4956	 &0.0776	&1.70E-03 \\
		& SubSim          &1.4501	&0.4371	 &0.0198	&2.49E-08 \\
		\bottomrule    
	\end{tabular}
	\label{t:optprice}
\end{table}

\begin{table}[t]
	\centering
	\caption{\textbf{Coefficient of variation (CV)}. A comparison of the CVs of the barrier option price as derived by each of the three methods (MCS, MLMC, SubSim) for four barrier levels against asset's volatility. }
	\begin{tabular}{llllll}
		\toprule
		& \multicolumn{5}{c}{Barriers}                                         \\
		&          & {[}60,140{]} & {[}70,130{]} & {[}80,120{]} & {[}90,110{]} \\
		\midrule
		Volatility ($\sigma$) & Method   & 
		&              &              &              \\
		\midrule
		0.05                  & Standard MCS & 0.0018       & 0.0016       & 0.0019       & 0.0045       \\
		& MLMC     & 0.0004       & 0.0004       & 0.0005       & 0.0024       \\
		& SubSim       & 0.0011       & 0.0013       & 0.0013       & 0.0031       \\
		0.10                  & Standard MCS & 0.0027       & 0.0026       & 0.0038       & 0.0080       \\
		& MLMC     & 0.0004       & 0.0006       & 0.0010       & 0.0077       \\
		& SubSim       & 0.0018       & 0.0019       & 0.0025       & 0.0059       \\
		0.15                  & Standard MCS & 0.0037       & 0.0040       & 0.0054       & 0.0177       \\
		& MLMC     & 0.0005       & 0.0008       & 0.0017       & 0.0229       \\
		& SubSim       & 0.0027       & 0.0026       & 0.0037       & 0.0092       \\
		0.20                  & Standard MCS & 0.0041       & 0.0053       & 0.0084       & 0.0444       \\
		& MLMC     & 0.0007       & 0.0013       & 0.0044       & 0.0598       \\
		& SubSim       & 0.0032       & 0.0039       & 0.0055       & 0.0156       \\
		0.25                  & Standard MCS & 0.0054       & 0.0066       & 0.0095       & 0.1122       \\
		& MLMC     & 0.0009       & 0.0020       & 0.0063       & 0.1623       \\
		& SubSim       & 0.0042       & 0.0053       & 0.0068       & 0.0219       \\
		0.30                  & Standard MCS & 0.0069       & 0.0089       & 0.0180       & 0.4069       \\
		& MLMC     & 0.0014       & 0.0053       & 0.0104       & 0.1992       \\
		& SubSim       & 0.0043       & 0.0061       & 0.0093       & 0.0347       \\
		0.35                  & Standard MCS & 0.0075       & 0.0099       & 0.0301       & 1.9758       \\
		& MLMC     & 0.0020       & 0.0041       & 0.0310       & 0.2169       \\
		& SubSim       & 0.0061       & 0.0072       & 0.0129       & 0.0652       \\
		0.40                  & Standard MCS & 0.0098       & 0.0126       & 0.0373       & 5.6981       \\
		& MLMC     & 0.0025       & 0.0057       & 0.0288       & 0.2257       \\
		& SubSim       & 0.0067       & 0.0088       & 0.0166       & 0.1047\\
		0.45                 & Standard MCS &  0.0087	& 0.0106	& 0.0254	&8.2893 \\
		& MLMC     & 0.0059	&0.0164	&0.0538	&0.2465      \\
		& SubSim       & 0.0077	&0.0128	&0.0217	&0.1808\\
		\bottomrule 
	\end{tabular}
	\label{t:COV1}
\end{table}
\end{document}